\documentclass[12pt]{iopart}

\pdfoutput=1

\usepackage{graphicx}
\usepackage{iopams}
\usepackage{url}

\newcommand{\ud}{\mathrm{d}}
\newcommand{\ui}{\mathrm{i}}

\begin{document}

\title[Financial instability from local market measures]{Financial instability from local market measures}

\author{Marco Bardoscia, Giacomo Livan, Matteo Marsili}
\address{Abdus Salam International Centre for Theoretical Physics, Strada Costiera 11, 34151 Trieste, Italy}
\eads{\mailto{marco.bardoscia@ictp.it}, \mailto{glivan@ictp.it}, \mailto{marsili@ictp.it}}

\begin{abstract}
We study the emergence of instabilities in a stylized model of a financial market, when different market actors calculate prices according to different (local) market measures. We derive typical properties for ensembles of large random markets using techniques borrowed from statistical mechanics of disordered systems. We show that, depending on the number of financial instruments available and on the heterogeneity of local measures, the market moves from an arbitrage-free phase to an unstable one, where the complexity of the market -- as measured by the diversity of financial instruments -- increases, and arbitrage opportunities arise. A sharp transition separates the two phases. Focusing on two different classes of local measures inspired by real markets strategies, we are able to analytically compute the critical lines, corroborating our findings with numerical simulations. 
\end{abstract}


\maketitle

\section{Introduction} 
\label{intro}

A huge number of financial instruments (bonds, futures, swaps, options, etc) are priced every day. Each of such instruments is generally priced according to models which approximate market dynamics and that need to be calibrated on market data \cite{Hull, BouchaudPotters}. In spite of the complexity of market dynamics, models used in pricing are often very simple: they depend on few parameters so as to allow for a very efficient and fast calibration on a small set of observed prices and rates\footnote{The number of model parameters is usually of the same order of magnitude of the number of required inputs from the market.}. This implies unavoidable approximations, both in the model and in the calibration, that are sometimes corrected in \emph{ad-hoc} manners\footnote{A typical example is the use of volatility ``smiles'' in the pricing of options according to Black-Scholes formula \cite{BouchaudPotters}, or the adoption of a replicating portfolio, that is a riskless strategy only in ideal markets.}.

Different institutions price different financial instruments according to different models, calibrating them against different data. As a consequence, not only the results are intrinsically approximated, but approximations are different for different financial instruments. Indeed even within the same financial institution, different models can be used to price different instruments. In financial jargon, this is a situation where each financial instrument is priced using a \emph{local} market measure, which may be specific of that instrument and/or of that institution. 

It has been argued \cite{Albanese1, Albanese2} that this practice can potentially lead to the emergence of arbitrage opportunities, \emph{i.e.}\ the possibility of making a riskless profit \cite{Pliska},  because the difference in the approximations used may give rise to a system of inconsistent prices. In ideal markets, such as those assumed as the basis of asset pricing theory (APT), which are infinitely liquid and complete, the presence of arbitrages would allow speculators to extract infinite profit, destabilizing markets. In real markets, the exploitation of arbitrages grants a finite payoff and this effectively ``closes'' the arbitrage opportunity (see e.g.\ \cite{MG} for a model of speculative activity that reproduces this mechanism). However, the trading strategies needed to exploit arbitrages may not exist or be feasible. In addition, in some cases misaligned prices may induce trading activity that enhances misalignment and that causes price instability (e.g. crashes)\footnote{If an asset is over-priced, the rational choice of selling it would drive price closer to its fundamental value. However, if speculators expect that others will buy because they expect the price to increase further, selling is not the best option. This is how financial bubbles self-sustain, amplifying an initial misalignment, until they burst. Furthermore, selling causes prices to decrease only when the total supply is fixed. If the supply can be expanded, speculation will not eliminate arbitrages. For example, in the recent bubble in credit derivatives,  misalignment in prices of Asset Backed Securities  (over-rating) was evident since at least early 2005 \cite{Rajan}. Still, the ``originate and distribute'' strategy has been widely practiced until mid 2007 by financial institutions, generating a skyrocketing expansion in the supply that finally led to the burst of the bubble.  The role of ``insurance portfolios'' during the 1987 crash (see \cite{BouchaudPotters} p.\ 179 for a short account) provides a further example of the relationship between arbitrages and instabilities. In the case, the departure of market dynamics from the Black-Scholes model accentuated the misalignment in prices, engendering cascades of sales leading to the crash. Notice, finally, that even when speculative activity removes arbitrages, works on Minority Games \cite{MG} suggests that increasing speculative activity causes excess volatility.}. 
This is why, in line with \cite{Albanese1,Albanese2}, we shall interchangeably talk about financial instabilities and arbitrage opportunities in what follows. 

Albanese \emph{et al.}\ \cite{Albanese1, Albanese2} propose that all instruments should be priced according to the same \emph{global} market measure, rather than with different {\em local measures}. For, in this case, the fundamental theorem of asset pricing \cite{Pliska, FTF} ensures that no arbitrage is possible. In \cite{Albanese1, Albanese2} it is further argued that the problems inherent in the use of local valuation techniques become particularly severe as the complexity of financial markets, both in terms of diversity of financial instruments and of volumes, increases.

In this paper, we address the issue of understanding what are the generic conditions under which the use of \emph{local} market measures can lead to market instability. We do this in a stylized setting where, however, we take on board the full  complexity of a market with a very large number of financial instruments. The use of tools of statistical mechanics of disordered systems \cite{Gardiner, Engel} allows us to characterize the typical behavior of ensembles of random markets, as in \cite{Marsili}. In this simplified setting, the difference in the pricing models is introduced as differences in the market measures used to price different financial instruments. This will allow us to focus on two key variables: the variability of risk measures across different assets and the number of assets. We shall show that, depending on these two variables, a market can be either found in an arbitrage free state or in one where arbitrage opportunities do exist. Arbitrages will be shown to typically arise when the number of different financial instruments is large. 

The rest of the paper is organized as follows: In Section \ref{Setup} we shall detail precisely the one-period economy framework, and, within such a context, we shall discuss global and local market measures, discussing their origin. We shall also provide a rigorous definition of arbitrage and introduce the arbitrage region volume, \emph{i.e.}\ the volume of the region in a given one-period economy's parameter space where arbitrage opportunities actually arise. In Section \ref{vol_computation} we shall provide a statistical mechanical approach to the evaluation of the arbitrage region volume, and in Section \ref{examples} we shall specialize the full general solution derived in Section \ref{vol_computation} to two meaningful examples. We shall then conclude the paper with some final remarks in Section \ref{concl}.

\section{Local and global measures in a one-period asset pricing model} 
\label{Setup}

Let us formalize this problem in the framework of a one-period economy. A one-period economy \cite{Pliska} is a setting characterized by two instants of time and $N$ different assets. At the initial time the price of asset $i$ is $p_i$. At the final time the world can be in any of $\omega = 1, \ldots, \Omega$ possible states and the amount paid by each asset depends on the realized state of the world; let $s_i^\omega$ be the amount paid at the final time by asset $i$ in the state of the world $\omega$. Therefore, $r_i^\omega=s_i^\omega-p_i$ is the corresponding return. An important result of APT\footnote{In the contest of APT it is customary to assume the existence of riskless asset (e.g.\ a bond) and to ``discount'' $p_i$ and $s_i^\omega$ with the corresponding quantities of the riskless asset. Our approach corresponds to considering discounted quantities from the beginning.} is that the market is arbitrage-free if and only if there exist (at least) one probability measure $q$ that (i) gives strictly positive weights to all states ($q^\omega>0, \forall \omega$), and (ii) such that the prices $p_i$ are the expected values of payoffs $s_i^\omega$ under $q$:
\begin{equation}
\label{price}
p_i=E_{q}[s_i] = \sum_{\omega=1}^\Omega q^\omega s_i^\omega, \qquad i=1,\ldots,N.
\end{equation}
Equation (\ref{price}) cannot be satisfied in general if $N > \Omega$ and, even in the case $N \leq \Omega$, it is not guaranteed that the solution is a probability distribution, \emph{i.e.}\ $q^\omega \geq 0$, $\forall \omega$. In particular, it is not guaranteed that the probability gives strictly positive weights to all states, meaning that for $N \leq \Omega$ there might still be arbitrage opportunities. 

A market is said to be complete if it is possible to exactly replicate every contingent claim that pays $s^\omega$ in state $\omega$ with a portfolio $\{z_i\}_{i=1}^N$ of assets, so that the return of the portfolio composed of one unit of the contingent claim plus the cover equals to zero in all the states of the world:
\begin{equation}
\label{replica_portfolio}
h^\omega \equiv s^\omega - p -\sum_{i=1}^N z_i(s_i^\omega - p_i) = 0 \qquad\forall\omega.
\end{equation}
It is clear that the market can be complete only if $N \geq \Omega$ and simultaneously complete and arbitrage-free only if $N = \Omega$; in the latter case the solution $q$ to (\ref{price}) is unique and, for any contingent claim that pays $s^\omega$ in state $\omega$, it is possible to compute its fair price as $p=E_q[s]$. In practice one is interested in a slightly weaker condition (perfect hedging):
\begin{equation}
\label{cnd0}
E_{q}[h]=E_{q}[h^2]=0,
\end{equation}
meaning that the contingent claim is replicated on average and with zero risk with respect to the probability measure $q$. The second equality in (\ref{cnd0}) implies
\begin{equation}
\label{condition}
h^\omega=0 \qquad \forall\omega:~q^\omega>0.
\end{equation}
In real life markets are incomplete. Therefore, provided that the market is arbitrage-free, the probability measure $q$ is not unique and it is not possible to find a riskless replicating portfolio such as the one in (\ref{replica_portfolio}). In other words, if $q$ has support on all states, then perfect hedging is not possible, in general, unless the market is complete. However, (\ref{cnd0}) can be satisfied if $q$ has support on a number $K$ of states smaller than $\Omega$. This would correspond, in the present context, to a \emph{local valuation} strategy, where the model used to price instrument $i$ is ``calibrated'' on $K$ different market outcomes. 
Under such local valuation scheme, one may think that, for each instrument, first a subset of assets $A$ with which to replicate a contingent claim is chosen; then, (\ref{price}) is solved for a subset of $K < \Omega$ of the states of the world, yielding a unique solution. Generically, this implies that the number $K$ of states used to ``calibrate'' the model equals the number of assets in the replicating portfolio: $K=|A|$. This is equivalent to taking $q^\omega=0$ on all states $\omega$ that are not among the $K$ chosen ones, as then (\ref{cnd0}) is still satisfied. 
This would not be a problem, as long as all contingent claims are priced in the same manner, \emph{i.e.}\ by employing the same set of assets $A$. This, as advocated in \cite{Albanese1, Albanese2}, would correspond to a \emph{global valuation} procedure.
However, if two instruments $i$ and $i^\prime$ are priced using different sets of stocks $A$ and $A^\prime$ and different sets of states, then this would imply two different \emph{local} measures $q$ and $q^\prime$ for the two instrument. Even if the two replicating portfolios used to price $i$ and $i^\prime$ would fulfill the conditions in (\ref{cnd0}) separately, \emph{i.e.}\ with respect to its own specific probability measure, the prices of instruments $s$ and $s^\prime$ would turn out to be inconsistent, and potentially lead to arbitrages.

There may be other ways in which a particular measure $q$ can be selected when pricing a given contingent claim. This may involve assumptions on the underlying processes, which may depend on a number of parameters. Again, these parameters may be set by ``calibrating'' the model in a number of market conditions or against observed prices of existing traded assets. In general, if a different procedure, model or approximation is used to compute the prices of different assets, this is equivalent to assuming a different measure $q_i$ for each asset $i$. We shall refer to this as to a \emph{local} probability measure in order distinguish it from the \emph{global} measure $q$ that should be used to price all assets coherently.

In this paper, we shall study the cases where either the measure $q_i^\omega$ is chosen to have its uniform support only on a subset of states (the subset is obviously different for different local measures), or when $q_i^\omega$ is drawn at random from some distribution, independently for each $i$. The key question we address is under what conditions the market will remain arbitrage free.

\subsection{Detecting arbitrage opportunities}

Given the previous discussion, let us consider a generic situation where $N$ assets are priced with different market measures $q_i$, $i=1,\ldots,N$. Thus, generalizing (\ref{price}), let us write
\begin{equation}
\label{price_local}
p_i = E_{q_i}[s_i] = \sum_{\omega=1}^\Omega q_i^\omega s_i^\omega, \qquad i = 1,\ldots,N
\end{equation}
where $q_i^\omega \geq 0, \forall \omega$, and of course $\sum_{\omega=1}^\Omega q_i^\omega = 1, \forall i$. 

Our goal is to check whether the existence of such $N$ different measures generates possible arbitrage opportunities. More specifically, suppose we have formed a portfolio with the $N$ assets available on the market by buying or borrowing an amount $z_i$ of asset $i$. Thus, the portfolio return in state $\omega$ will read
\begin{equation}
\label{portfolio_return}
r_z^\omega = \sum_{i=1}^N z_i \left (s_i^\omega - p_i  \right).
\end{equation} 
Now, an arbitrage opportunity emerges whenever one can devise a portfolio yielding a non-negative return in all possible states of the world, \emph{i.e.}\ $r_z^\omega \geq 0, \forall \omega$. One way to measure ``how many'' arbitrage opportunities are there in our one-period set-up, is to compute the following volume in the $N$-dimensional space of portfolio weights:
\begin{equation} 
\label{arbitrage_vol}
V = \int_{-\infty}^{+\infty} \ud \mathbf{z} \prod_{\omega=1}^\Omega \Theta \left ( \sum_{i=1}^N z_i ( s_i^\omega - p_i ) \right ),
\end{equation}
where $\ud \mathbf{z} = \prod_{i=1}^N \ud z_i$, and $\Theta(\ldots)$ denotes Heaviside's step function. We shall refer to the quantity in (\ref{arbitrage_vol}) as to the arbitrage region volume (or, more simply, arbitrage volume) throughout the rest of this paper.

The arbitrage region volume defined in (\ref{arbitrage_vol}) is delimited by the $\Omega$ constraints
\begin{equation} 
\label{constr}
\sum_{i=1}^N z_i ( s_i^\omega - p_i ) \geq 0
\end{equation}
in the $N$-dimensional space $z_1, \ldots, z_N$. Each constraint defines a $(N-1$)-dimensional hypersurface passing through the origin of such space. As a consequence, depending on the mutual compatibility of the constraints, the arbitrage region must be either unbounded (we shall say infinite in the following) or equal to zero. Already at this level, geometrical intuition suggests that, if the number of constraints is smaller than or equal to the number of dimensions, \emph{i.e.}\ if $N \geq \Omega$, a region where arbitrage opportunities arise should exist. This suggests that the interesting scaling for the issue at stake is the one where $N$ is proportional to $\Omega$. Let us then introduce the rescaled variable
\begin{equation}
\label{ratio}
n=\frac{N}{\Omega}.
\end{equation}
In the following we shall focus on the limit $N,\Omega\to\infty$ with $n$ finite: in such a framework, our previous observation can be rephrased by stating that the arbitrage volume will be infinite for $n \geq 1$, while for $n < 1$ it will either be zero or infinite depending on other specificities we shall discuss in the next sections.

We consider an ensemble of random realizations of this problem, where the payoffs $s_i^\omega$ are independent and identically distributed (i.i.d.)\ random variables drawn from a Gaussian distribution with zero mean. Already at this level we do not expect any result to depend on the payoffs' variance (provided it is different from zero), since any rescaling of $s_i^\omega$, hence of $p_i$, can be absorbed by rescaling the weights $z_i$ accordingly (see (\ref{arbitrage_vol})). Thus, we will fix such variance to be equal to one. On the other hand, for the sake of generality we shall not make, for the moment, any distributional assumption on the probabilities $q_i^\omega$.

Before moving to the analytical computation of the arbitrage volume, let us mention that, for each instance, it can be evaluated numerically by means of linear programming\footnote{Actually, the function $f$ to be minimized is not important at all, since one is only interested in understanding if a region of the space of the assets such that the linear constraints (\ref{constr}) are satisfied exists. As already discussed, the corresponding linear programming problem will be either unbounded, meaning that the coordinates of the optimal solution diverge and that the arbitrage volume is infinite, or it will have the trivial solution $z_i = 0 \; \forall i$, meaning that the constraints cannot be simultaneously satisfied in a region of the space of non-zero volume, and thus that the arbitrage volume is zero. However, since all the regions of the asset space delimited by the constraints (\ref{constr}) are open polytopes with a vertex in the origin, depending on the orientation of the polytope and on the function to minimize, the linear programming problem can admit the trivial solution, even if the volume is infinite. However, this problem can be easily sidestepped by checking whether also the linear programming problem for the function $-f$ only admits the trivial solution.}. We shall indeed compare our theoretical prediction for typical properties in the limit $N,\Omega\to\infty$ with the behavior of single instances for finite $N$ and $\Omega$.

The next section discusses the calculation of the arbitrage volume with tools of statistical mechanics of disordered systems. The calculation is standard \cite{Gardiner,Engel}, but, for the sake of completeness, we provide its main conceptual steps nonetheless. The reader who is familiar with these techniques, or who is not interested in the full technical details, may skip it and go directly to the next section.

\section{Computing the arbitrage region volume for large $N$}
\label{vol_computation}

The volume $V$ in (\ref{arbitrage_vol}) clearly depends on the specific realizations of prices $s_i^\omega$ and probabilities $q_i^\omega$. Still, the question we are interested in has the flavor of a threshold phenomenon, and the thresholds at which the volume diverges for two different realizations are expected to be close when $N$ is very large. Put differently, it is expected that different realizations ``typically'' behave in a similar manner for large $N$. Typical behavior is generally related to quantities that are \emph{self-averaging}, \emph{i.e.}\ that satisfy a law of large numbers (or a concentration principle), and the experience in statistical mechanics of disordered systems teaches us that one has to look for \emph{extensive} quantities, \emph{i.e.}\ quantities proportional to the size $N$ of the system. In our case, it is reasonable to think of the arbitrage region, when shrinking from being unbounded to zero volume, as a box of volume $d^N$ for some typical scale $d$. Thus, the extensive quantity we are looking for is $\log V$, and we anticipate here that, for any realization, we shall find 
\begin{equation}
\label{logV}
v \equiv \lim_{N\to\infty} \frac{1}{N} E_{s,q} [\log V] = \lim_{N\to\infty} \frac{1}{N}\log V
\end{equation}
where $E_{s,q} [\ldots]$ stands for the average on different realizations of prices and probabilities. Please notice the change of notation: in (\ref{logV}), and in the following, $E_q[\ldots]$ denotes the average over the probability density describing the random variables $q_i^\omega$, whereas in (\ref{price}) and (\ref{cnd0}) it denotes the average over the non-random set of probabilities $q^\omega$. Computing averages of logarithms is a technical difficulty we shall circumvent by using the replica trick:
\begin{equation}
\label{replica_trick}
E_{s,q} [\log V]=\lim_{r \to 0} \frac{1}{r}\log E_{s,q} [V^r].
\end{equation}

The volume is calculated a la Gardiner \cite{Gardiner}. Our first step to explicitly compute the averaged arbitrage volume will be to consider $r$ replicas of the volume in (\ref{arbitrage_vol}):
\begin{equation}
\label{replica_vol}
V^r = \prod_{a=1}^r \int_{-\infty}^{+\infty} \ud \mathbf{z}_a \prod_{\omega=1}^\Omega \Theta \left ( \sum_{i=1}^N z_{ia} y_i^\omega \right ),
\end{equation}
where we have posed $y_i^\omega = s_i^\omega - p_i$ and $\ud \mathbf{z}_a = \prod_{i=1}^N \ud z_{ia}$. Making use of the integral representation of Heaviside's $\Theta$ function, it is possible to explicitly compute the average of $V^r$ in (\ref{replica_vol}) with respect to the probability distributions of the random variables $s_i^\omega$ and $q_i^\omega$, \emph{i.e.}\ the variables $y_i^\omega$. The result reads
\begin{eqnarray}
\label{avg_replica_vol}
\fl E_{s,q} [ V^rÊ] = E_y [V^r] &=& \int_{-\infty}^{+\infty} \left ( \prod_{a=1}^r \ud \mathbf{z}_a \right) \int_{-\infty}^{+\infty} \left ( \prod_{a=1}^r \frac{\ud \mathbf{k}_a}{(2\pi)^\Omega} \right) \int_0^{+\infty} \left ( \prod_{a=1}^r \ud \mathbf{x}_a \right) \\ \nonumber
& \cdot &  \exp \left ( - \frac{1}{2} \sum_{a,b=1}^r \sum_{i=1}^N \sum_{\omega,\omega^\prime=1}^\Omega k_a^\omega k_b^{\omega^\prime} z_{ia} z_{ib} Y_{\omega \omega^\prime} + \ui \sum_{a=1}^r \sum_{\omega=1}^\Omega k_a^\omega x_a^\omega \right ),
\end{eqnarray}
where a few new objects have been introduced. Let us comment on those. First, in the previous expression we have posed $\ud \mathbf{k}_a = \prod_{\omega=1}^\Omega \ud k_a^\omega$ (with an analogous expression for $\ud \mathbf{x}_a$), where the integration variables $k_a^\omega$ and $x_a^\omega$ arise from the aforementioned integral representation of Heaviside's function. On the other, $Y_{\omega \omega^\prime}$ represents the covariance matrix of the random variables $y_i^\omega$:
\begin{equation}
\label{cov_matrix}
Y_{\omega \omega^\prime} = E_{s,q} [ y_i^\omega y_i^{\omega^\prime} ] = E_y [ y_i^\omega y_i^{\omega^\prime} ].
\end{equation}
Let us remind the reader that no distributional assumptions on the probabilities $q_i^\omega$ have been made yet. However, the functional form in (\ref{avg_replica_vol}) emerges through the following approximation. Let us pose for a moment $\Gamma_i^\omega = \sum_{a=1}^r k_a^\omega z_{ia}$:
\begin{eqnarray}
\label{avg_approx}
E_y \left [ \exp \left ( - \ui \sum_{\omega=1}^\Omega \Gamma_i^\omega y_i^\omega \right ) \right ] &\sim& \prod_{i=1}^N \left ( 1 - \frac{1}{2} \sum_{\omega, \omega^\prime = 1}^\Omega \Gamma_i^\omega \Gamma_i^{\omega^\prime} E_y [y_i^\omega y_i^{\omega^\prime}] \right ) \\ \nonumber
&\sim& \exp \left ( \frac{1}{2} \sum_{i=1}^N \sum_{\omega, \omega^\prime = 1}^\Omega \Gamma_i^\omega \Gamma_i^{\omega^\prime} Y_{\omega \omega^\prime} \right ),
\end{eqnarray}
where we also used $E_y [y_i^\omega y_j^{\omega^\prime}] = \delta_{ij} Y_{\omega \omega^\prime}$ and $E_y[y_i^\omega] = 0$ (which is a straightforward consequence of $E_s[s_i^\omega] = 0$).

In what follows, we shall assume the covariance matrix in (\ref{cov_matrix}) to be defined according to the following structure (an assumption satisfied by the examples we shall considered later):
\begin{equation}
\label{cov_matrix_struct}
Y_{\omega \omega^\prime} = \delta_{\omega \omega^\prime} + \frac{y}{\Omega}.
\end{equation}
With this position, after taking the limit for the number of replicas $r$ going to zero in (\ref{replica_trick}), the averaged logarithm of the arbitrage volume, is (we refer the interested reader to \ref{calculations} for a detailed derivation of this equation) 
\begin{equation}
	v = \max_{\chi, \hat{\chi}, \gamma, \hat{\gamma}, \hat{\sigma}, \phi} \tilde{g}(\chi, \hat{\chi}, \gamma, \hat{\gamma}, \hat{\sigma}, \phi)
\end{equation}
where
\begin{eqnarray} 
\label{htilde}
\fl \tilde{g} = \frac{\chi}{2} \left ( \hat{\chi} - \hat{\sigma}^2 + \hat{\gamma}^2 \right ) + \frac{\hat{\chi} \phi}{2} - \gamma \hat{\gamma} &+& E_t \left [ \log \int_{-\infty}^{+\infty} \ud z \exp \left ( - \frac{\hat{\chi}}{2} z^2 + \hat{\sigma} t z \right ) \right ] \\ \nonumber
&+& \frac{1}{n} E_w \left [ \log \frac{1}{2} \mathrm{Erfc} \left ( \frac{\sqrt{\phi} \ w + \gamma \sqrt{n y }}{\sqrt{2 \chi}} \right ) \right ],
\end{eqnarray}
where $t$ and $w$ are two standard Gaussian random variables, whereas $E_t[\ldots]$ and $E_w[\ldots]$ denote the expectations with respect to the corresponding probability measures. From the previous expression one can immediately see that any dependence of the arbitrage volume on the distributional structure of the prices $s_i^\omega$ and the probabilities $q_i^\omega$ is actually encoded in the parameter $y$. In the case in which $y < 0$, we rewrite (\ref{htilde}) for convenience by introducing
\begin{eqnarray} 
\label{paramslist2}
\lambda = \ui \gamma \ , \ \ \ \hat{\lambda} = \ui \hat{\gamma},
\end{eqnarray}
so that (\ref{htilde}) reads
\begin{eqnarray} 
\label{htilde2}
\fl \tilde{g} = \frac{\chi}{2} \left ( \hat{\chi} - \hat{\sigma}^2 - \hat{\lambda}^2 \right ) + \frac{\hat{\chi} \phi}{2} + \lambda \hat{\lambda} &+& E_t \left [ \log \int_{-\infty}^{+\infty} \ud z \exp \left ( - \frac{\hat{\chi}}{2} z^2 + \hat{\sigma} t z \right ) \right ] \\ \nonumber
&+& \frac{1}{n} E_w \left [ \log \frac{1}{2} \mathrm{Erfc} \left ( \frac{\sqrt{\phi} \ w + \lambda \sqrt{n |y|}}{\sqrt{2 \chi}} \right ) \right ].
\end{eqnarray}
In the following the calculations will be detailed for the case $y < 0$ and straightforwardly generalized to the complementary case.

The limit where the arbitrage volume shrinks to zero is found as the limit where the distance between two solutions goes to zero. So, since 
\begin{equation} 
\label{chi}
\chi = \frac{1}{2N} \sum_{i=1}^N (z_{ia} - z_{ib})^2
\end{equation}
is actually the distance between two replicas, as one can easily check from the definition of $\chi$ in (\ref{paramslist}), such a limit is found for $\chi \rightarrow 0$. Moreover, under this limit the volume is reasonably expected to scale as $V \sim \chi^N = \exp(N \log \chi)$. Thus, the leading contribution to $v$ must be of order $\log \chi$. Also, by computing the saddle point equations on $\tilde{g}$ in (\ref{htilde}) (which essentially implements the $N \rightarrow \infty$ limit) it can be shown that $\phi$ and $\lambda$ reach finite limits when $\chi \rightarrow 0$ (see also \cite{Marsili} for a similar calculation), whereas the remaining parameters in (\ref{htilde}) behave as $\hat{\chi} = k / \chi$, $\hat{\sigma} = s / \chi$ and $\hat{\lambda} = \lambda / \chi$. In the light of the above considerations on the leading contribution to $v$, we must make sure that its terms which ``unphysically'' scale as $1 / \chi$ are canceled when $\chi \rightarrow 0$, and this is obtained by computing
\begin{eqnarray} 
\label{htildalim}
\tilde{v} &=& \lim_{\chi \rightarrow 0} \chi v \\ \nonumber 
&=& \frac{1}{2} \left ( k \phi - s^2 + \lambda^2 \right ) + E_t \left [\max \left ( - \frac{k}{2} z^2 + t s z \right ) \right ] - \frac{\phi}{2n} I_2(w_0), 
\end{eqnarray}
where $I_n(w_0) = E_w \left [ (w + w_0)^n \Theta(w + w_0) \right ]$ and $w_0 = \lambda \sqrt{n |y| /\phi}$. Performing the saddle point analysis on $\tilde{v}$ in (\ref{htildalim}), and taking the derivatives of $\tilde{v}$ with respect to $s$ and $k$ one has
\begin{eqnarray} 
\label{sp}
\frac{\partial \tilde{v}}{\partial s} &=& E_t [t z^*] - s = 0 \\ \nonumber
\frac{\partial \tilde{v}}{\partial k} &=& \frac{\phi}{2} - \frac{1}{2} E_t [(z^*)^2] = 0,
\end{eqnarray}
where $z^* = \mathrm{arg\, max} \left ( - \frac{k}{2} z^2 + t s z \right ) = st/k$. The previous equations give $k = 1$ and $\phi = s^2$, so that $\tilde{v}$ can be rewritten as
\begin{equation} 
\label{htildesp}
\tilde{v} = \frac{s^2 + \lambda^2}{2} - \frac{s^2}{2n} I_2(w_0) = \frac{s^2}{2} \left [ 1 + \xi^2 - \frac{1}{n} I_2(\xi \sqrt{n |y|}) \right ],
\end{equation}
where we have posed $\xi = w_0 / (\sqrt{n |y|})$. Eventually, the saddle point equations on $s$ and $\xi$ read
\begin{eqnarray} 
\label{sp2}
\frac{\partial \tilde{v}}{\partial s} = 0 \qquad &\Rightarrow& \qquad 1 + \xi^2 - \frac{1}{n} I_2(\xi \sqrt{n |y|}) = 0 \\ \nonumber
\frac{\partial \tilde{v}}{\partial \xi} = 0 \qquad &\Rightarrow& \qquad - \xi + \sqrt{\frac{|y|}{n}} I_1(\xi \sqrt{n |y|}) = 0.
\end{eqnarray}

For the cases in which $y > 0$ one can proceed in complete analogy to the previous case. When doing so, (\ref{htildesp}) is rewritten as
\begin{equation} \label{htildesp2}
\tilde{v} = \frac{s^2 - \gamma^2}{2} - \frac{s^2}{2n} I_2(w_0) = \frac{s^2}{2} \left [ 1 - \xi^2 - \frac{1}{n} I_2(\xi \sqrt{n y}) \right ],
\end{equation}
where $\hat{\gamma} = \gamma / \chi$, $w_0 = \gamma \sqrt{n y/\phi}$, $\xi = w_0/(\sqrt{n y})$. The saddle point equations in this case read
\begin{eqnarray} \label{sp3}
\frac{\partial \tilde{v}}{\partial s} = 0 \qquad &\Rightarrow& \qquad 1 - \xi^2 - \frac{1}{n} I_2(\xi \sqrt{n y}) = 0 \\ \nonumber
\frac{\partial \tilde{v}}{\partial \xi} = 0 \qquad &\Rightarrow& \qquad \xi + \sqrt{\frac{y}{n}} I_1(\xi \sqrt{n y}) = 0.
\end{eqnarray}
Jointly solving the sets of equations (\ref{sp2}) or (\ref{sp3}) provides the relation between $n$ and $y$ (or any parameter contained in the definition of $y$) on the boundary of the critical region where the arbitrage volume becomes equal to zero. In the next section we shall discuss some examples by specifying the covariance matrix (\ref{cov_matrix}).

\section{Examples of local measures}
\label{examples}

Now, in order to practically use (\ref{sp2}) and (\ref{sp3}) we essentially need to make a distributional assumption on the probabilities $q_i^\omega$ and consequently compute $y$ via (\ref{cov_matrix_struct}). The very general structure of the covariance matrix $Y_{\omega \omega^\prime}$ in (\ref{cov_matrix}) is the following:
\begin{eqnarray} \label{Ystruct}
Y_{\omega \omega^\prime} &=& E_y [y_i^\omega y_i^{\omega^\prime}] = E_s [s_i^\omega s_i^{\omega^\prime}] + E_{s,q} [p_i^2] - E_{s,q} [p_i s_i^\omega] - E_{s,q} [p_i s_i^{\omega^\prime}] \\ \nonumber
&=& \delta_{\omega \omega^\prime} + \sum_{\omega^{\prime \prime} = 1}^\Omega E_q [(q_i^{\omega^{\prime \prime}})^2] - E_q [q_i^\omega] - E_q [q_i^{\omega^\prime}] ,
\end{eqnarray}
which actually matches the structure we had assumed in (\ref{cov_matrix_struct}), with $y$ equal to $\Omega$ times the second term in (\ref{Ystruct}). It is worth noting that in all the practical cases that we will discuss, $y$ is (at most) of order one (see (\ref{kappaprob}) and (\ref{deltavar})) ensuring the (semi-)positive definiteness of the covariance matrix $Y_{\omega \omega^\prime}$.

\subsection{Hedging on a subset of states}

\begin{figure}
	\centering
	\includegraphics[scale=0.7]{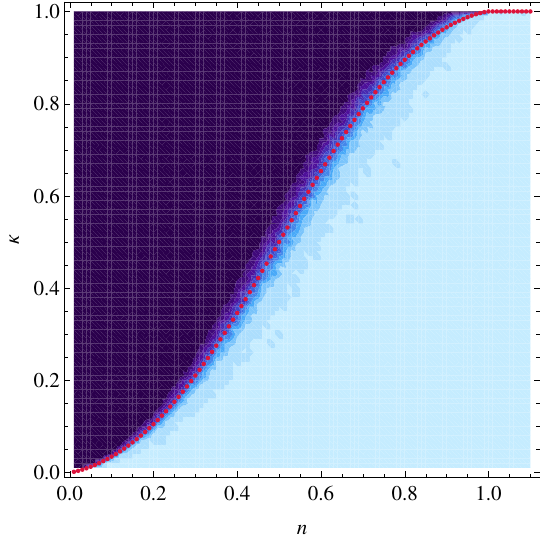}
	\caption{Contour plot of the arbitrage volume as a function of $n$ and $\kappa$ for $q_i^\omega$ chosen as in (\ref{qK}). Average over $100$ realizations of $q_i^\omega$ and $s_i^\omega$ and $N=100$. The region on the right (light blue) corresponds to infinite volume, while the region on the left (dark blue) corresponds to zero volume. A sharp transition occurs at the boundary between the two regions. The critical line is completely recovered by the analytical solution of (\ref{sp2}) and (\ref{sp3}) (red dots).}
	\label{fig:subset}
\end{figure}

The first case we wish to address is the one, already qualitatively discussed in Section \ref{intro}, where each instrument $i$ in the market is hedged considering only an instrument-dependent subset $\Omega_i$ made of $K \leq \Omega$ market states. We assume probabilities to be uniform over such subsets:
\begin{equation} \label{qK}
q_i^\omega = \left \{ \begin{array}{rl}
1/K & \mathrm{if} \; \omega \in \Omega_i \\
0 & \mathrm{if} \; \omega \notin \Omega_i.
\end{array} \right.
\end{equation} 
In the thermodynamic limit one has that
\begin{equation} \label{qK}
q_i^\omega = \left \{ \begin{array}{rl}
1/K & \mathrm{with \ probability} \; K / \Omega \\
0 & \mathrm{with \ probability} \; 1 - K / \Omega,
\end{array} \right.
\end{equation} 
and it is then immediate to compute the following quantities:
\begin{eqnarray} \label{kappaprob}
\mathbb{E}_q [q_i^\omega] &=& \frac{1}{\Omega} \\ \nonumber
\mathbb{E}_q [(q_i^\omega)^2] &=& \frac{1}{K \Omega},
\end{eqnarray}
so that the covariance matrix (\ref{Ystruct}) is straightforwardly computed and, $y$ reads
\begin{equation} 
\label{kappaconst}
y = \sqrt{\frac{1}{\kappa} - 2},
\end{equation}
where $\kappa = K / \Omega$. When $\kappa \in (0,1/2]$ the set of saddle point equations (\ref{sp3}) can be used to calculate the critical line, whereas when $\kappa \in (1/2,1]$ one needs to use the (\ref{sp2}) with $|y| = \sqrt{2 - 1/ \kappa}$. In Figure \ref{fig:subset} we compare the obtained critical line with numerical simulations (see Section \ref{Setup}). We show the numerically evaluated arbitrage volume averaged over different realization of prices and probabilities as a function of $\kappa$ and $n$. It can be immediately noticed that for $n \geq 1$ the volume is always infinite, consistently with the observations made in Section \ref{intro}. At fixed $\kappa$ one has that, for a sufficiently small density of instruments $n$, the arbitrage volume is zero; when increasing $n$ the system crosses a sharp transition and enters the phase in which the arbitrage volume is infinite. The analytically calculated critical line in the plane $(n, \kappa)$ closely matches the sharp transition observed in the numerical simulations.

\subsection{Perturbation of uniform probabilities}

\begin{figure}
	\centering
	\includegraphics[scale=0.7]{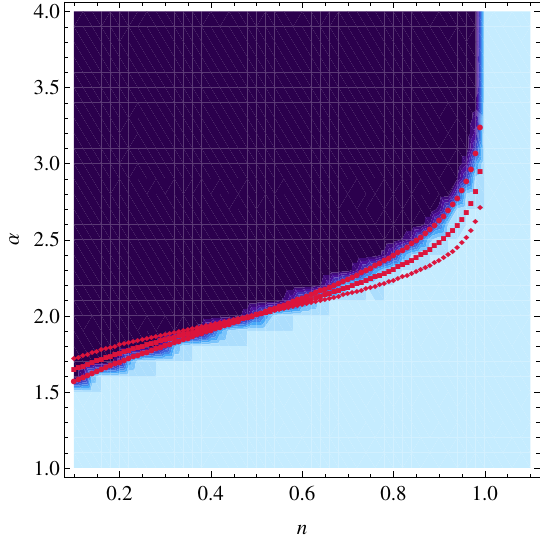}
	\caption{Contour plot of the arbitrage volume as a function of $n$ and $\alpha$ for $q_i^\omega$ chosen as in (\ref{perturb}) with $\Delta=1$. Average over $100$ realizations of $q_i^\omega$ and $s_i^\omega$ and $N=200$. The same comments made in the caption of Figure \ref{fig:subset} apply here. The critical lines for $N=10^3$ (red squares) and $N=10^4$ (red diamonds) are also shown, signaling the fact that, in the thermodynamic limit, the critical line becomes a sharp edge at $n = 1$ and $\alpha = 2$.}
	\label{fig:prtb}
\end{figure}

Let us now discuss the case
\begin{equation} 
\label{perturb}
q_i^\omega = \frac{1}{\Omega} + \Delta_i^\omega,
\end{equation}
where $\Delta_i^\omega$ is a Gaussian random variable with zero mean and variance equal to $\Delta / \Omega^\alpha$. This choice amounts to assume that all instruments are hedged with a flat probability distribution ($1/\Omega$, $\forall \omega$) plus an instrument-dependent ``perturbation'' $\Delta_i^\omega$. Let us denote as $E_\Delta[\ldots]$ the expectation value with respect to the multivariate Gaussian distribution of the $\Delta_i^\omega$, and let us enforce the constraint $\sum_{\omega=1}^\Omega \Delta_i^\omega = 0$ in order to ensure the normalization (but not necessarily the positive definiteness) of the $q_i^\omega$ in (\ref{perturb}). It is then immediate to verify that $E_\Delta[\Delta_i^\omega \delta(\sum_{\omega^\prime=1}^\Omega \Delta_i^{\omega^\prime})] = 0$. On the other hand, it is a little more tricky to compute the following quantity:
\begin{eqnarray} 
\label{deltavar}
\fl E_\Delta \left [ (\Delta_i^\omega)^2 \delta \left (\sum_{\omega^\prime=1}^\Omega \Delta_i^{\omega^\prime} \right ) \right ] = \\ \nonumber
\fl - 2 \frac{\partial}{\partial \lambda_\omega} \left \{ \log \int_{-\infty}^{+\infty} \left [ \prod_{\omega^\prime = 1}^\Omega \frac{\ud \Delta_i^{\omega^\prime}}{\sqrt{2 \pi \Delta / \Omega^\alpha}} \exp \left ( - \frac{\lambda_\omega}{2} (\Delta_i^{\omega^\prime})^2 \right ) \right] \delta \left ( \sum_{\omega^\prime = 1}^\Omega \Delta_i^{\omega^\prime} \right ) \right \} \Bigg |_{\lambda_\omega = \frac{\Omega^\alpha}{\Delta}} \\ \nonumber
\fl = \frac{\Delta}{\Omega^\alpha} \left ( 1 - \frac{1}{\Omega} \right ) \sim \frac{\Delta}{\Omega^\alpha}.
\end{eqnarray}
These results can be used to compute the covariance matrix $Y_{\omega \omega^\prime}$ in (\ref{cov_matrix}) and $y$. For $\Delta \leq \Omega^{\alpha -2}$ we have
\begin{equation}
\label{A_perturb}
|y| = \sqrt{1 - \frac{\Delta}{\Omega^{\alpha-2}}}.
\end{equation}
Having in mind a situation in which $\Delta$ is of order one while $\Omega \rightarrow \infty$, (\ref{A_perturb}) effectively holds for $\alpha \geq 2$. In this case, when taking $\Omega \rightarrow \infty$ as required by the thermodynamic limit, one has of course $|y| \rightarrow 1$, so that the saddle point equations (\ref{sp2}), and therefore the position of the critical line, do not depend neither on $\alpha$ nor on $\Delta$. From (\ref{sp2}) one has that a transition occurs at $n = 1$ in this case. On the other hand, when $\alpha < 2$ one has $y = \sqrt{\Delta / \Omega^{\alpha-2} - 1} \rightarrow \infty$ for $\Omega \rightarrow \infty$, and the arbitrage volume diverges. As a result, the arbitrage volume depends on $\Delta$, \emph{i.e.}\ on the typical size of the fluctuation of the probability measure $q$, only for $\alpha = 2$. The phase diagram is characterized by a sharp corner at the intersection of the two lines $\alpha = 2$ and $n = 1$. However, from (\ref{A_perturb}) one expects the convergence to the thermodynamic limit to be slow for $2 < \alpha < 3$; for this reason, in Figure \ref{fig:prtb} we compare the numerically evaluated arbitrage volume with the critical line obtained by solving (\ref{sp2}) in which the proper finite-size value of $\Omega$ has been plugged. Two additional critical lines calculated using higher values of $N$ (which reflect the corresponding finite-size values of $\Omega$) are also shown, clearly indicating that the critical line becomes a sharp edge in the thermodynamic limit.

\begin{figure}
	\centering
	\includegraphics[scale=0.7]{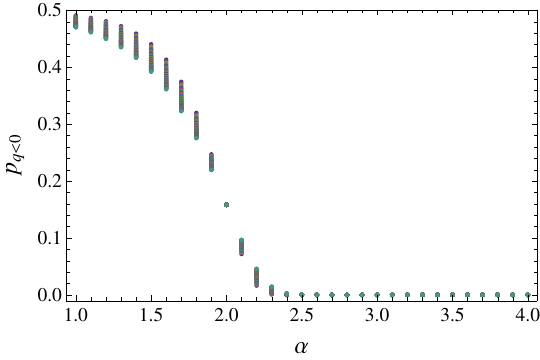}
	\includegraphics[scale=0.7]{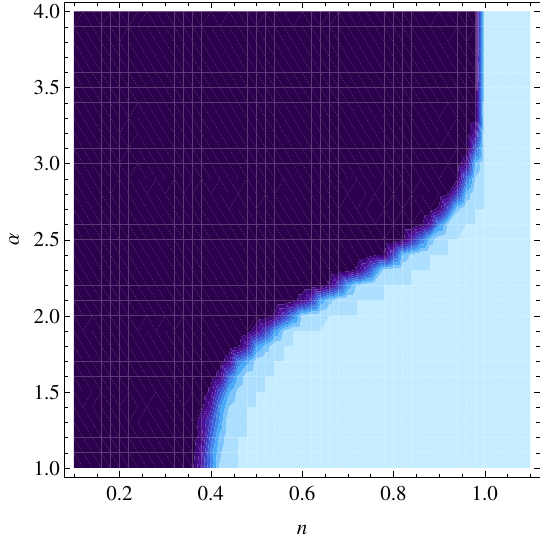}
	\caption{Left panel: fraction $p_{q<0}$ of negative probabilities $q_i^\omega$ with respect to $\alpha$. For each value of $\alpha$ we show the values of $p_{q<0}$ corresponding to different values of $n$ spanning the interval $[0.1, 1.1]$. Right panel: contour plot as in Figure \ref{fig:prtb}, but with hard constraint on $q_i^\omega$. As expected from the left panel, the difference with the case in Figure \ref{fig:prtb} is relevant only for $\alpha < 2.5$.}
	\label{fig:prtb_pneg}
\end{figure}

As previously recalled, the choice in (\ref{perturb}) does not prevent from the possibility of having negative values for the $q_i^\omega$, which of course would prevent from interpreting them as probabilities. From the left panel of Figure \ref{fig:prtb_pneg} we see that the fraction of negative probabilities $p_{q<0}$ depends significantly only on $\alpha$, going from a region where it is zero to a region where it becomes macroscopic. The change occurs approximately at $\alpha \simeq 2.5$. A further check is provided by numerically evaluating the arbitrage volume with $q_i^\omega$ chosen as in (\ref{perturb}), but implementing a hard constraint on their sign; this means that all the negative $q_i^\omega$ are set to zero, and the remaining ones are normalized to one. From the right panel of Figure \ref{fig:prtb_pneg} we see that, as expected, the arbitrage value is significantly different from the one obtained without the hard constraint on $q_i^\omega$ only for $\alpha < 2.5$.

\begin{figure}
	\centering
	\includegraphics[scale=0.7]{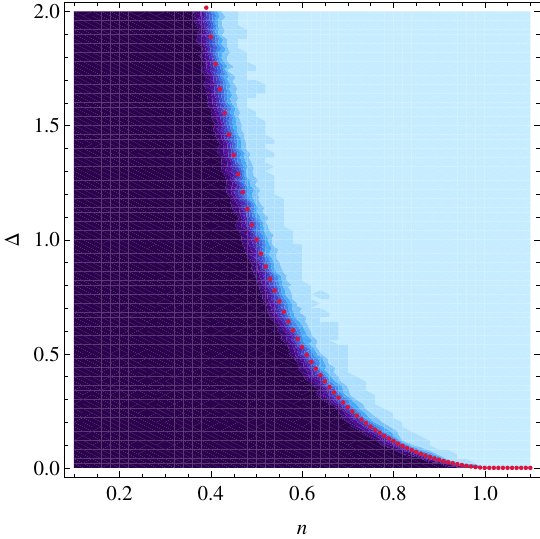}
	\includegraphics[scale=0.7]{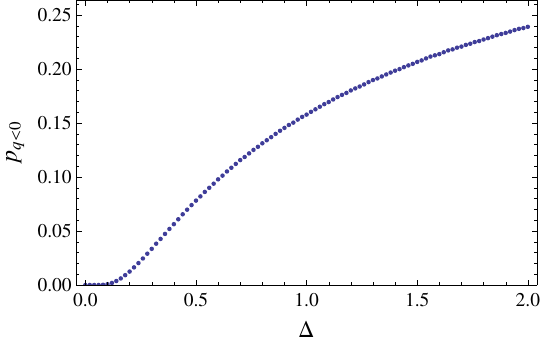}
	\caption{Left panel: contour plot of the arbitrage volume as a function of $n$ and $\Delta$ for $q_i^\omega$ chosen as in (\ref{perturb}) with $\alpha=2$. Average over $100$ realizations of $q_i^\omega$ and $s_i^\omega$ and $N=200$. The same comments made in the caption of Figure \ref{fig:subset} apply here. Right panel: fraction $p_{q<0}$ of negative probabilities $q_i^\omega$ with respect to $\Delta$.}
	\label{fig:delta}
\end{figure}

The case in which $\alpha = 2$ can be studied with (\ref{sp2}) and $|y| = \sqrt{1 - \Delta}$ if $\Delta \geq 1$, and with (\ref{sp3}) and $y = \sqrt{\Delta - 1}$ if $\Delta < 1$. In the left panel of Figure \ref{fig:delta} we show that, also in this case, the analytically calculated critical line is in excellent agreement with the boundary of the transition from numerical simulations. If $\alpha = 2$, the fraction of negative probabilities $p_{q<0}$ does not depend on $\Omega$ and, as shown in the right panel of Figure \ref{fig:delta}, it is equal to zero only for small values of $\Delta$.

\section{Conclusions} 
\label{concl}

In summary, we have shown that the use of local market measures leads to the emergence of arbitrages in a simplified model of a financial market. Although the setup is very idealized, our results corroborate the claims of  \cite{Albanese1, Albanese2} that call for the use of global market measures for asset pricing. Indeed, as argued in \cite{Albanese1, Albanese2}, we find that instabilities arise precisely when the complexity of financial markets, as measured by the diversity of financial instruments, increases. In the case of the hedging on a subset of states, arbitrage opportunities arise for small values of $\kappa$,  \emph{i.e.}\ when the overlap between different local measures is smaller. In the case of perturbation of uniform probabilities that happens for small values of $\alpha$, \emph{i.e.}\ when the deviations are larger. These findings support those of \cite{Marsili,Hommes} that suggest an inverse relation between financial complexity and systemic stability, and provides an additional element of caution against the unfettered development of financial markets. 

While it is hard to relate our conclusions to realistic settings of specific markets, we believe that the emergence of instabilities in an idealized framework of competitive and perfectly liquid markets is an important proof of concept, calling for the need to consider global market measures for asset pricing. In real markets, arbitrages are often temporary and are quickly removed by speculative activity (see discussion in the Introduction). Here we find that an increased complexity generates inconsistent prices, and in their turn, arbitrage opportunities attract speculators. 
It is tempting to speculate that the fact that the recent surge in complexity of derivative markets has been accompanied by a parallel escalation in the activity of hedge funds might not be a coincidence. This suggests that including elements of global valuation in regulatory frameworks might temper the dynamics of financial markets and promote financial stability.

\appendix
\section{Derivation of the average volume}
\label{calculations}

Our first step is (\ref{avg_replica_vol}) for the averaged replicated volume. After posing $Y_{\omega \omega^\prime} = \delta_{\omega \omega^\prime} + y / \Omega$, as in (\ref{cov_matrix_struct}), let us perform a Hubbard-Stratonovich transformation to decouple the replica indices in (\ref{avg_replica_vol}). When doing so one obtains
\begin{eqnarray} 
\label{avgrvol2}
\fl E_{s,q} [ V^rÊ] &=& \int_{-\infty}^{+\infty} \left ( \prod_{a=1}^r \ud \mathbf{z}_a \right) \int_0^{+\infty} \left ( \prod_{a=1}^r \ud \mathbf{x}_a \right) \int_{-\infty}^{+\infty} \left ( \prod_{a=1}^r \frac{\ud \mathbf{k}_a}{(2\pi)^\Omega} \right) \\ \nonumber 
\fl & \cdot & \int_{-\infty}^{+\infty} \left ( \prod_{i=1}^N \frac{\ud t_i}{\sqrt{2\pi}} \right ) \exp \left ( - \frac{1}{2} \sum_{i=1}^N t_i^2 - \frac{1}{2} \sum_{a,b=1}^r \sum_{i=1}^N \sum_{\omega=1}^\Omega k_a^\omega k_b^\omega z_{ia} z_{ib} \right ) \\ \nonumber
\fl & \cdot & \exp \left ( \ui \sqrt{\frac{y}{\Omega}} \sum_{i=1}^N t_i \sum_{a=1}^r \sum_{\omega=1}^\Omega k_a^\omega z_{ia} + \ui \sum_{a=1}^r \sum_{\omega=1}^\Omega k_a^\omega x_a^\omega \right ).
\end{eqnarray}
Let us now introduce the following order parameters
\begin{equation} \label{ordparams}
\phi_{ab} = \frac{1}{N} \sum_{i=1}^N z_{ia} z_{ib} \ , \ \ \ \gamma_a = \frac{1}{N} \sum_{i=1}^N t_i z_{ia},
\end{equation}
and let us enforce their definition within (\ref{avgrvol2}) by means of the identity
\begin{eqnarray} \label{identity}
1 &=& \int_{-\infty}^{+\infty} \ud \phi_{ab} \ N \delta \left ( N \phi_{ab} - \sum_{i=1}^N z_{ia} z_{ib} \right ) \\ \nonumber
&=& \int_{-\infty}^{+\infty} \int_{-\infty}^{+\infty} \frac{\ud \phi_{ab} \ud \hat{\phi}_{ab}}{4 \pi \ui / N} 
\exp \left ( - \frac{N}{2} \sum_{a,b=1}^r \phi_{ab} \hat{\phi}_{ab} + \frac{1}{2} \sum_{a,b=1}^N \sum_{i=1}^N \hat{\phi}_{ab} z_{ia} z_{ib} \right ),
\end{eqnarray}
plus analogous equations for the $\gamma_a$. With the previous definitions the averaged volume in (\ref{avgrvol2}) can be written as
\begin{eqnarray} \label{avgrvol3}
E_{s,q} [ V^rÊ] &=& \int_{-\infty}^{+\infty} \left (\prod_{a,b=1}^r \frac{\ud \phi_{ab} \ud \hat{\phi}_{ab}}{4 \pi \ui / N} \right ) \int_{-\infty}^{+\infty} \left (\prod_{a=1}^r \frac{\ud \gamma_a \ud \hat{\gamma}_a}{2 \pi \ui / N} \right ) \\ \nonumber 
&\cdot& \exp \left ( N g \left (\{\phi_{ab}\}, \{\hat{\phi}_{ab}\}, \{\gamma_a\}, \{\hat{\gamma_a}\} \right ) \right ),
\end{eqnarray}
where $g = g_1 + g_2 + g_3$ and
\begin{eqnarray} \label{gfunc}
g_1 \left (\{\phi_{ab}\}, \{\hat{\phi}_{ab}\}, \{\gamma_a\}, \{\hat{\gamma_a}\} \right ) = - \frac{1}{2} \sum_{a,b=1}^r \phi_{ab} \hat{\phi}_{ab} - \sum_{a=1}^r \gamma_a \hat{\gamma}_a \\ \nonumber
g_2 \left (\{\hat{\phi}_{ab}\}, \{\hat{\gamma_a}\} \right ) = \log \int_{-\infty}^{+\infty} \frac{\ud t}{\sqrt{2 \pi}} \int_{-\infty}^{+\infty} \left ( \prod_{a=1}^r \ud z_a \right ) \\ \nonumber
\phantom{g_2 \left (\{\hat{\phi}_{ab}\}, \{\hat{\gamma_a}\} \right )} \cdot \;\; \exp \left (- \frac{1}{2} t^2 + \frac{1}{2} \sum_{a,b=1}^r \hat{\phi}_{ab} z_a z_b + t \sum_{a=1}^r \hat{\gamma}_a z_a \right ) \\ \nonumber
g_3 \left (\{\phi_{ab}\}, \{\gamma_a\} \right ) = \frac{\Omega}{N} \log \int_{-\infty}^{+\infty} \left ( \prod_{a=1}^r \frac{\ud k_a}{2\pi} \right ) \int_{-\infty}^{+\infty} \left ( \prod_{a=1}^r \ud x_a \right ) \\ \nonumber
\phantom{g_3 \left (\{\phi_{ab}\}, \{\gamma_a\} \right )} \cdot \;\; \exp \left ( -\frac{1}{2} \sum_{a,b=1}^r k_a k_b \phi_{ab} + \ui \sqrt{\frac{y}{\Omega}} \sum_{a=1}^r \gamma_a k_a + \ui \sum_{a=1}^r k_a x_a \right ).
\end{eqnarray}

Under a replica symmetric ansatz, we can write the order parameters as
\begin{equation} \label{RSA}
\phi_{ab} = (\Phi - \phi) \delta_{ab} + \phi \ , \ \ \ \gamma_a = \gamma,
\end{equation}
with analogous definitions for $\hat{\phi}_{ab}$ and $\hat{\gamma}_a$. So, by also posing $n = N / \Omega$, as in (\ref{ratio}), one can write the following relations for the functions in (\ref{gfunc}) when taking the limit $r \rightarrow 0$:
\begin{eqnarray} \label{gfunc2}
\tilde{g}_1 &=& \lim_{r \rightarrow 0} \frac{g_1}{r} = - \frac{1}{2} (\Phi \hat{\Phi} - \phi \hat{\phi}) - \gamma \hat{\gamma} \\ \nonumber
\tilde{g}_2 &=& \lim_{r \rightarrow 0} \frac{g_2}{r} = E_t \left [ \log \int_{-\infty}^{+\infty} \ud z \exp \left ( \frac{1}{2} (\hat{\Phi} - \hat{\phi}) z^2 + t z \sqrt{\hat{\phi} + \hat{\gamma}^2} \right) \right ] \\ \nonumber
\tilde{g}_3 &=& \lim_{r \rightarrow 0} \frac{g_3}{r} = \frac{1}{n} E_w \left [ \log \frac{1}{2} \mathrm{Erfc} \left  ( \frac{\sqrt{\phi} \ w + \gamma \sqrt{n y}}{\sqrt{2(\Phi - \phi)}} \right ) \right ],
\end{eqnarray} 
where $t$ and $w$ are Gaussian variables with zero mean and unit standard deviation. In these expressions we directly computed the $r \rightarrow 0$ limit on the functions $g_1$, $g_2$ and $g_3$ which actually appear as the argument of the exponential function in (\ref{avgrvol3}). This is because, according to (\ref{logV}) and (\ref{replica_trick}), one has $\lim_{N \rightarrow \infty} N^{-1} \log V = \lim_{N \rightarrow \infty} \lim_{r \rightarrow 0} N^{-1} r^{-1} E_{s,q}[\log V^r]$. So, keeping in mind that the limit for $N \rightarrow \infty$ will eventually be taken, reducing the integral in (\ref{avgrvol3}) to its saddle point approximation, one can compute the $r \rightarrow 0$ limit directly on the intensive function of the order parameters $g = g_1 + g_2 + g_3$ in (\ref{avgrvol3}), as we did.

By introducing the following set of parameters
\begin{eqnarray} 
\label{paramslist}
\chi = \Phi - \phi \ , \ \ \ \hat{\chi} = \hat{\phi} - \hat{\Phi} \ , \ \ \ \hat{\sigma} = \sqrt{\hat{\phi} + \hat{\gamma}^2} \ , 
\end{eqnarray}
we eventually get to (\ref{htilde}) noting from (\ref{logV}) that $\tilde{g} = \tilde{g}_1 + \tilde{g}_2 + \tilde{g}_3$.

\section*{References}


\begin{thebibliography}{99}

\bibitem{Hull} 
Hull J, 
2008 
\emph{Options, futures, and other derivatives} 
(Upper Saddle River NJ: Pearson).

\bibitem{BouchaudPotters} 
Bouchaud J-P and Potters M, 
2000 
\emph{Theory of Financial Risk and Derivative Pricing} 
(Cambridge: Cambridge University Press). 

\bibitem{Albanese1} 
Albanese C, Gimonet G and White S, 
\emph{Towards a Global Valuation Model}, 
2010 \emph{Risk Magazine} May issue 72.

\bibitem{Albanese2} 
Albanese C, Bellaj T, Gimonet G and Pietronero G, 
\emph{Coherent Global Market Simulations and Securitization Measures for Counterparty Credit Risk}, 
2011 \emph{Quant. Fin.} {\bf 11} 1.

\bibitem{Pliska}
Pliska S R, 
1997 
\emph{Introduction to Mathematical Finance} 
(Oxford, Blackwell).

\bibitem{MG} 
Challet, D, Marsili, M, Zhang, Y C, 
2005 
\emph{Minority Games} 
(Oxford: Oxford University Press).

\bibitem{Rajan} 
Rajan, G R, 
\emph{Has financial development made the world riskier?}, 
2005 NBER Working Paper Series, No. 11728, November 2005 
\url{http://ideas.repec.org/p/nbr/nberwo/11728.html}.

\bibitem{FTF} 
de Finetti B, 
\emph{Sul Significato Soggettivo della Probabilit\`{a}}, 
1931 \emph{Fundamenta Mathematic\ae} \textbf{17} 298.

\bibitem{Gardiner}
Gardiner E, 
\emph{The space of interactions in neural network models}, 
1988 \emph{J. Phys. A: Math. Gen.} \textbf{21} 257.

\bibitem{Engel} 
Engel A, 
2001 
\emph{Statistical mechnics of learning} 
(Cambridge: Cambridge University Press).

\bibitem{Marsili}
Marsili M, 
\emph{Complexity and financial stability in a large random economy}, 
2009 SSRN: \url{http://ssrn.com/abstract=1415971}.

\bibitem{Hommes} 
Brock W A, Hommes C H and Wagener F O O, 
\emph{More hedging instruments may destabilize markets}, 
2008 CeNDEF Working paper 08-04, University of Amsterdam 
\url{http://ideas.repec.org/p/ams/ndfwpp/08-04.html}.


\end{thebibliography}
\end{document}